\documentclass[useAMS,usenatbib,usegraphicx,a4]{article}
\usepackage{times}
\usepackage{url}
\usepackage{epsfig}
\usepackage[flushleft]{threeparttable}
\usepackage{amssymb}
\usepackage{amsmath}
\usepackage{pdfpages}
\usepackage{array}
\usepackage{comment}
\usepackage{hyperref} 
\usepackage{graphicx}
\usepackage{longtable}
\usepackage{ltablex}
\usepackage{threeparttable}
\usepackage{multirow}
\usepackage{rotating}
\usepackage[export]{adjustbox}
\usepackage{aas_macros}
\usepackage{authblk}
\newcolumntype{H}{>{\setbox0=\hbox\bgroup}c<{\egroup}@{}}
\usepackage[sort&compress,super,comma]{natbib}

\begin{document}

\title{Universality in the Random Walk Structure Function of Luminous Quasi-Stellar Objects}

\author[1]{\href{mailto:ji-jia.tang@anu.edu.au}{Ji-Jia Tang}}
\author[1,2]{\href{mailto:christian.wolf@anu.edu.au}{Christian Wolf}}
\author[3]{John Tonry}

\affil[1]{Research School of Astronomy and Astrophysics, Australian National University, Cotter Road Weston Creek, ACT 2611, Australia}
\affil[2]{Centre for Gravitational Astrophysics, Australian National University, Building 38 Science Road, Acton, ACT 2601, Australia}
\affil[3]{Institute for Astronomy, University of Hawaii, 2680 Woodlawn Drive, Honolulu, HI 96822-1897, U.S.A.}

\date{draft \today}
\maketitle

\begin{abstract}
Rapidly growing black holes are surrounded by accretion disks that make them the brightest objects in the Universe. Their brightness is known to be variable, but the causes of this are not implied by simple disk models and still debated. Due to the small size of accretion disks and their great distance, there are no resolved images addressing the puzzle. In this work, we study the dependence of their variability on luminosity, wavelength and orbital/thermal timescale. We use over 5,000 of the most luminous such objects with light curves of almost nightly cadence from $>5$ years of observations by the NASA/ATLAS project, which provides 2 billion magnitude pairs for a structure function analysis. When time is expressed in units of orbital or thermal time scale in thin-disk models, we find a universal structure function, independent of luminosity and wavelength, supporting the model of magneto-rotational instabilities as a main cause. Over a $>1$~dex range in time, the fractional variability amplitude follows $\log (A/A_0) \simeq 1/2 \times \log (\Delta t/t_{\rm th})$. Deviations from the universality may hold clues on the structure and orientation of disks.
\end{abstract}

\vspace{0.05cm}

Every massive galaxy in the Universe contains a supermassive black hole at its centre; their origin and growth is one of the biggest mysteries in astronomy. Most black holes appear dark and are revealed only by their gravity \cite{KH13}. During growth spurts, however, matter from surrounding areas falls towards the black hole. Its gravitational energy is turned via heat into radiation, which highlights the black hole as a "QSO" (Quasi-Stellar Object). Fast-growing black holes are the most luminous objects in the Universe. At UV-optical wavelengths, QSOs can outshine 1,000 large galaxies \cite{Ri06,Wo18}.

The key for turning gravitational energy into heat is matter flowing through an accretion disk \cite{PR72}. Matter orbiting the black hole drifts inwards as high friction in the disk dissipates orbital energy. It is now clear that Keplerian disks with weak magnetic fields are prone to magneto-rotational instability (MRI)\cite{BH91}. MRI is a fundamental instability converting differential rotation into turbulent fluid motions, feeding an energy cascade, and explaining the anomalously high viscosity needed for luminous QSO disks\cite{Ba03}.

Accretion disks are an elegant explanation for the high luminosity and compactness of QSOs. The standard model of optically thick and geometrically thin disks predicts their temperature profiles and emission\cite{SS73}. As they are too small to be spatially resolved by current facilities, we seek clues about their properties from the variability of their emission, which is ubiquitous on all timescales. While MRI must be one source of short-term fluctuations, another must be the rapidly fluctuating heating from the X-ray corona near the black hole\cite{Cl92, Pe93, Ca07}. The ``lamppost'' model exploits the latter and relates the sizes of accretion disks to time lags between light curves of different passbands probing regions of different temperature and radius in the disk\cite{Pe93,Ca07,Sh14, Fa16, Ji17, Fa18, Yu20}.

Monitoring of QSO samples has revealed that fluctuations can be described by a ‘damped random walk’ model \cite{Ke09, Mc10, Wa19}: brightness changes tend to be larger when the time delay between any two measurements is longer. However, this model provides only a behavioural analogy and no physical insight. Mathematically, the light curve of an object can be statistically described by a variability "structure function" (SF): the typical brightness difference is expressed as a function of time interval \cite{Hu92, Ko16}. The SFs of QSOs are characteristically different from other types of variable objects, so that QSOs can even be identified in a time-domain survey by the SF alone \cite{Pa11}.

Measurements of the SF agree on two trends: fluctuations on a fixed timescale get stronger in less luminous QSOs and in bluer passbands\cite{VB04, Mc10, Mo14, Ca17, Li18, Su21}. Thus, the SF can be parametrised as 
\begin{equation}\label{SF_basic}
 \log A = B_0 + \gamma \times \log \Delta t
            + B_L \times \log L
            + B_\lambda \times \log \lambda  ~,
\end{equation} 
where amplitudes $A$ of fractional flux changes are in terms of magnitude, and $L$ is the measured luminosity of the QSO. Both the time separation between the two measurements, $\Delta t$, and the wavelength of observed light, $\lambda$, are expressed in the QSO rest-frame with cosmological redshift and time dilation removed. Outside of the random-walk regime, the amplitudes may be suppressed at shortest timescales \cite{Mu11, Ka15a, Sm18, Ta20, St22} for reasons that are not yet established. They tend to level off above a damping or decorrelation timescale\cite{Mc10, Bu21, St22}, which has been shown to be driven by the thermal timescale of the accretion disk\cite{Bu21}.

Past works agree that $B_L<0$ and $B_\lambda<0$ but disagree on the strength of the trends. 
Some complicating factor may not be controlled for in existing studies. 
The SF slope $\gamma = d\log A/d\log \Delta t$ at fixed $(L,\lambda)$ is in the range\cite{VB04, Mo14, Ko16, Li18} of $0.2$ to $0.6$. $B_L = d\log A/d\log L$ at fixed $(\Delta t,\lambda)$ describes the dependence on disk luminosity (hence size) and is reported in the range\cite{VB04, Mc10, Mo14, Li18, Su21} of $-0.33$ to $-0.16$. Finally, $B_\lambda = d\log A/d\log \lambda$ at fixed $(\Delta t,L)$ describes the trend with rest-frame wavelength and thus location of the fluctuating source within the disk; measurements range \cite{Mc10, Mo14, Li18} from $-0.75$ to $-0.48$, although observations of non-linear relations also include slope values outside this range. 

Here, we present a new analysis of the variability structure function of the $\sim 5,000$ brightest QSOs in the sky using light curves with unprecedented sampling. The sample has a median black-hole mass of $\sim 2$~billion solar masses according to an analysis of spectra from the Sloan Digital Sky Survey (SDSS)\cite{Ra20}. The light curves are from the 0.5-metre NASA/ATLAS telescope (Asteroid Terrestrial-impact Last Alert System\cite{To18a}), which was built to find dangerous asteroids that might threaten the Earth. ATLAS is designed to image the entire sky seen from its location in Hawaii four times every night, weather permitting. With up to 1,200 nights per QSO per band, we get $\sim 2.1$ billion magnitude pairs to constrain the QSO SF. This is one of the best data sets for studying QSO variability before the new Vera C. Rubin Telescope operates.

\section*{Results}

The new light curves include observations from 2015 to 2021 in two passbands, cyan and orange (centred on 533 and 679 nm, respectively). We limit the sample to redshift $0.5<z<3.5$ ($z<2.4$ in the cyan band), where QSOs are so luminous that their host galaxies will not affect the measurements. We reject radio-loud QSOs, where variability may be enhanced by shocks in jets and mask the pure accretion disk behaviour \cite{He76}. We also reject gravitationally lensed QSOs and QSOs with close neighbours as seen by the ESA Gaia mission\cite{gaia21} to avoid blended light curves.

Previous work typically fitted a global model for the SF, such as Equation~\ref{SF_basic}, to a whole data set. The richness of this new data set offers the alternative to split the data into subsets without incurring the penalty of noisy statistics. We first split our data into twenty bins in $\Delta t$ and fit the trends with luminosity and wavelength separately for each bin, using the equation
\begin{equation}
\log (A (\Delta t)) = B_{0,\Delta t} 
                + B_{L,\Delta t} \times \log L
                + B_{\lambda,\Delta t} \times \log \lambda  ~.
\label{eq:SF_dt}
\end{equation} 
Thus, we get twenty different estimates for $B_L$ and $B_\lambda$, independently for each $\Delta t$ bin. As we can now study the trends on different timescales, we may see different regimes of behaviour. We use the measured variability on timescales of 1 day to estimate the noise in the observations. At the long end, window effects from the finite length of the light curves may cause noise in the parameters \cite{Ko17}. We restrict the wavelength range to longer than 130 nm (avoiding flux from the strong Ly-$\alpha$ emission line) and shorter than 300 nm (avoiding a previously seen upturn in variability\cite{VB04, Mc10, Ca17}, which is currently unexplained and discussed further below). We use two types of statistical analysis, simple bin averages, and a computationally more intense bootstrap analysis. 

\begin{figure}
\begin{center}
\includegraphics[width=0.8\columnwidth]{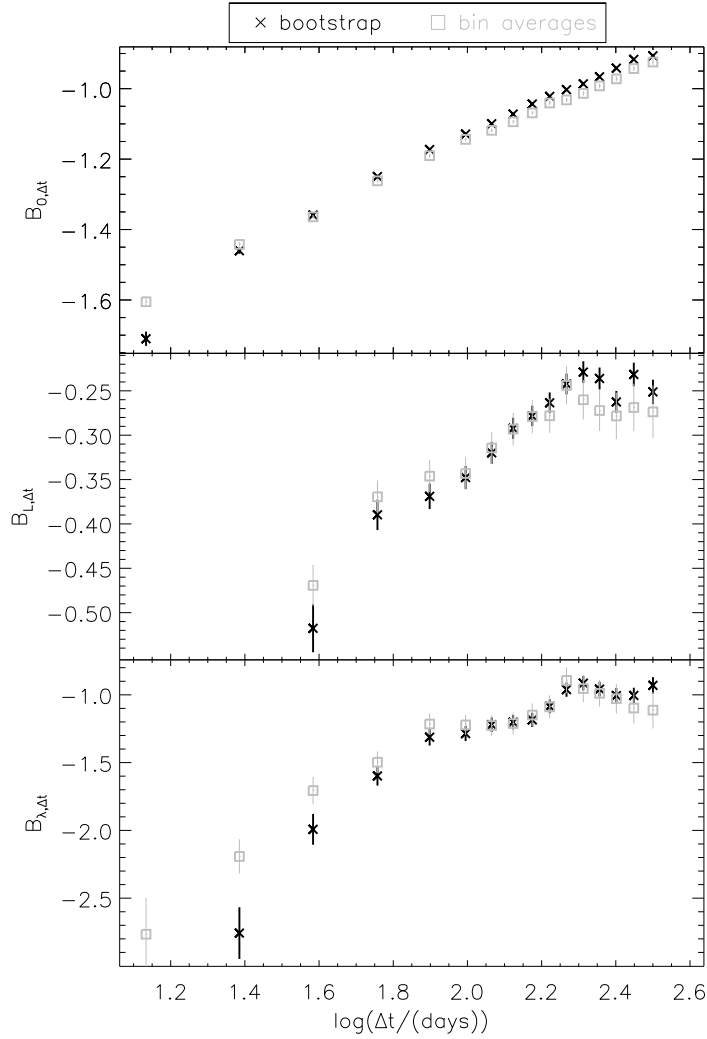}
\caption{Coefficients of the structure function from Equation~\ref{eq:sf_l_lambda_fit} in narrow time intervals, comparing bootstrap results and simple bin averages. 
This figure contains the result of 5315 QSOs. The bootstrap error bars are the standard deviation of the distribution while the error bars of the simple bin averages are the standard error of the median (SEMED). Data points and error bars are based on $\sim 2$ billion magnitude pairs from 5315 QSOs.
}
\label{boot}
\end{center}
\end{figure}

\begin{figure}
\begin{center}
\includegraphics[width=0.65\columnwidth]{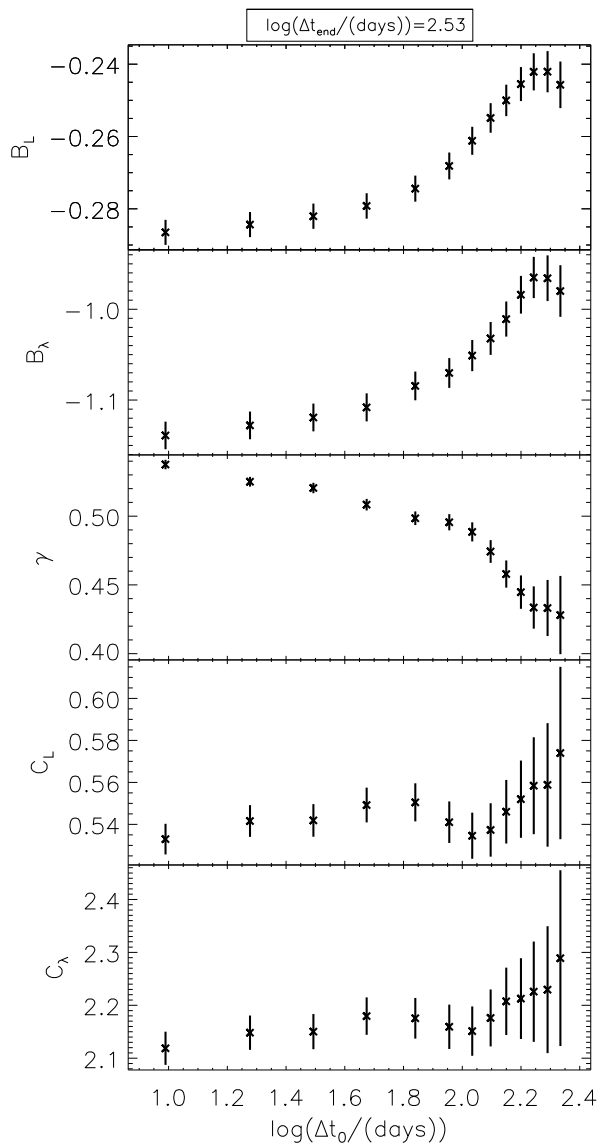}
\caption{Global coefficient estimates from Equation~\ref{eq:sf_coes_dt_fit} (bootstrap case). $\Delta t_0$ is the short end of the fitting window, while the long end of the window ($\Delta t_{\rm end}$) is fixed at 337~days. The crosses and their error bars are the means and the standard deviations of the bootstrap distributions. When short timescales are included, the estimates drift, as non-constant behaviour is included into the fit. Restricting the window to the longest timescales, shrinks the data set and makes the fits noisier (data points and error bars are based on $\sim 2$ billion magnitude pairs from 5315 QSOs).
}
\label{mean_parameters}
\end{center}
\end{figure}

Figure \ref{boot} shows the resulting parameters for a range in $\Delta t$ from 10 to 337 days. Both methods of analysis produce consistent estimates. The estimates for $B_L$ and $B_\lambda$ show a strong trend with $\Delta t$. It is immediately clear that a global fit will produce estimates for $B_L$ and $B_\lambda$ that depend on the fitting range in $\Delta t$ and on the distribution of pairs in the range, and thus on subtleties of survey cadence. We suggest this fact explains the variety of estimates obtained in the past. We demonstrate how results depend on the range of behaviour forced into a global fit in Figure \ref{mean_parameters}, where fiducial global estimates for $B_L$ and $B_\lambda$ are obtained by choosing different global fitting ranges for $\Delta t$. 

Finally, we improve the global fit by restricting it to the random-walk regime. Given that our sample comprises the most luminous known QSOs, we expect damping timescales well in excess of a year, which do not affect our analysis. Having controlled for $\Delta t$ by fitting the trends in each bin separately (Figure \ref{boot}), we reveal {\it two} regimes in behaviour: at short $\Delta t$ both $B_L$ and $B_\lambda$ drift with $\Delta t$, whereas at longer $\Delta t$ they both remain constant. Thus, at longer $\Delta t$ the choice of fitting range will have little impact. Here, we find fit parameters of $\gamma \approx 1/2$, $B_L \approx -1/4$ and $B_\lambda \approx -1$. Past estimates of $B_L$ cover values from a lower range\cite{Mc10,Su21} of $-0.33$ to $-0.29$, depending on the chosen parametrisation, to an upper limit\cite{Mo14} of $-0.16$ and include values very close\cite{VB04, Li18} to $-1/4$.

\subsection*{Changing the clock for a universal structure function}

So far, $B_L$ and $B_\lambda$ have described how variability amplitudes behave at fixed $\Delta t$. We now look at the timescales, on which disks of different $L$, observed at different $\lambda$, show a fixed amplitude $A$ and thus potentially a universal structure function if their slopes $\gamma$ are the same as well. We find these timescales to be parametrised as:
\begin{eqnarray}
    \label{eq:CL}
  C_L       = d\log \Delta t (A)/d\log L        & = & -B_L/\gamma = 0.539\pm0.004 \\
   \label{eq:Clambda}
  C_\lambda = d\log \Delta t (A)/d\log \lambda  & = & -B_\lambda/\gamma = 2.418\pm0.023  ~.
\end{eqnarray}

The first study in this direction empirically found $C_L=0.47\pm 0.06$ and argued that the fractional optical variability per disk orbital or thermal timescale is likely constant\cite{Ke13}. This argument is based on the well-known fact that the predicted temperature profile in thin disks\cite{SS73}, $T(R) \propto R^{-3/4}$, sets a mean radius for the origin of radiation and thus a mean orbital timescale $t_{\rm orb}$ of the emitting surface, given $\lambda$ and a size, and thus $L$, of the disk\cite{Mc10}. Note, that for a fixed viscosity of the disk material, the thermal timescale $t_{\rm th}$ is proportional to the orbital timescale and we get\cite{Ke13}:
\begin{equation}\label{eq:tth_obs}
 \log t_{\rm orb} \propto \log t_{\rm th} \propto 1/2 \log L + 2 \log \lambda ~.
\end{equation} 

One later study\cite{Ca17} reported $C_L\simeq 0.41\pm 0.04$. Our result shows that the concept of a constant fractional variability per orbital or thermal timescale\cite{Ke13} is not only consistent with measurements of $C_L$ but also our $C_\lambda$. Thus, we rephrase the whole SF from Equation~\ref{SF_basic} with time in new units, choosing those of thermal timescale given its role for the damping timescale\cite{Bu21}:
\begin{equation}\label{A_ttherm}
   \log (A/A_0) = \gamma_{\rm th} \log (\Delta t/t_{\rm th}) ~.
\end{equation}
This form is expected for fluctuations arising from MRI, where the equations of motion scale with the orbital timescale (and thus the thermal timescale)\cite{Ba03,FKR02}. It also applies in the corona-heated accretion-disk reprocessing (CHAR) model\cite{Su20}, which is motivated by the concern that a lamppost will not create sufficient amplitude by heating alone. The CHAR model proposes that the corona couples directly to the magnetic field in the disk and drives stronger fluctuations than expected from a lamppost corona alone.

We determine $t_{\rm th}$ for each QSO and passband given its $(L,\lambda)$. Figure \ref{SF_dttherm} shows the resulting SF expressed both in natural time and with a $t_{\rm th}$-scaled clock. Here, we increased the time resolution but reduced noise by grouping QSOs into subsamples split by $(L,\lambda)$. The wide $(L,\lambda)$-range of our sample causes a wide range of amplitudes at fixed $\Delta t$ (left panel), but at fixed $\Delta t/t_{\rm th}$ (right panel) we find indeed a tight range, supporting the concept of a universal structure function with a constant slope and intercept once expressed in suitable units of time. The relation holds over more than 1 dex in $\Delta t/t_{\rm th}$, limited by the length of our light curves at long timescales. At short timescales, the amplitude suppression relative to the random walk remains. The CHAR model actually predicts a suppression with a break timescale proportional to the thermal timescale, which would imply a fixed break timescale in Figure \ref{SF_dttherm}, right panel. However, we see a spread of timescales such that the break depends more strongly on luminosity than the thermal timescale, which will be investigated in future work. For our final global fit of the SF, we avoid the suppressed regime by using only data at $\log A>-1.4$. We find a bootstrap slope with natural clock time of $\gamma = 0.503\pm 0.001$ and with time in units of thermal timescale of $\gamma_{\rm th}=0.510\pm 0.002$, extremely consistent with the $\gamma =1/2$ of a random walk (Table~\ref{tab:sf_coes} summarises all fit parameters).

\subsection*{Outlook}

The transition from a regular clock to an orbital or a thermal time-scaled clock has standardised the random walk of QSOs as the wide dispersion observed among  objects with a variety of luminosities and at a variety of restframe wavelengths has disappeared. Instead, the SF of QSOs appears now as an intrinsically tight relation. 

A universal SF opens a window to characterise QSOs in more detail by deviations from the mean relation using future higher-precision measurements. Such deviations are expected to arise from any effect that varies the orbital and thermal timescale estimates. These effects include the dependence of the observed luminosity on disk orientation and black-hole spin\cite{Ca18}, which are not independently known for individual objects; the true temperature profile of the accretion disk, which is still debated as it may change from $T\propto R^{-3/4}$ in an inner part driven by viscosity to $T\propto R^{-1/2}$ in an outer part dominated by irradiation from an X-ray corona acting as a lamppost\cite{FKR02}; and the actual interior structure of the disk, which may well change from thin to slim disks as the accretion rate approaches the Eddington rate\cite{Ab88,FKR02,Sa11}.

\begin{figure}
\begin{center}
\includegraphics[width=\textwidth]{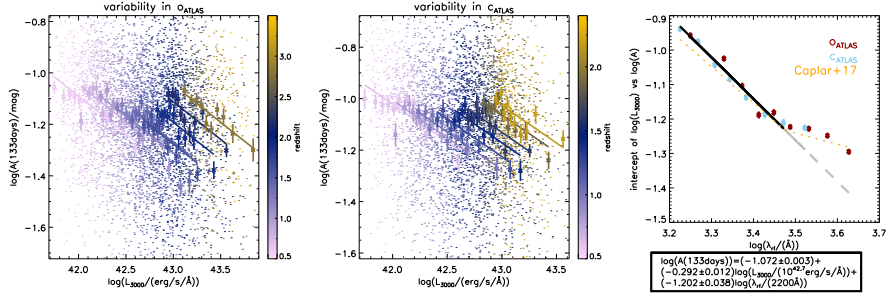}
\caption{One example time separation bin, $\Delta t_{\rm rest}=[125;141]$ days. The 5,315 individual QSOs are points, while large symbols and their error bars are bootstrap mean and standard deviation of the variability amplitudes in bins of luminosity $L_{3000}$ and redshift $z$, lines are fits to bootstrap mean values (left: orange passband; centre: cyan passband). Right: intercepts and errors of the fits at fixed $L_{3000}$ for different rest-frame wavelengths, where either passband is seen to sample the same underlying behaviour. Solid lines are fit ranges, dashed lines are extrapolations shown for clarity. Note an upwards deviation at $\lambda > 300$~nm, which is consistent with the dotted line from one past work \cite{Ca17} and also seen in other works \cite{VB04, Yu22}. 
}
\label{SF_dt9}
\end{center}
\end{figure}

Deviations from a simple picture are generally evident at wavelengths of $\lambda > 300$~nm. Here, we measure an upturn in $A$, relative to a trend of $B_\lambda \approx -1$, which has been seen before \cite{VB04, Mc10, Ca17, Yu22} (see also Figure~\ref{SF_dt9}, right panel). It suggests that fluctuations in the outer, cooler parts of QSO accretion disks are to be enhanced by an additional mechanism, which has not yet been explained. Also, we excluded $\lambda < 130$~nm, as we cannot disentangle light from the disk continuum and light from the broad Ly $\alpha$ line. The hotter inner disks are harder to study as a wide swath of spectrum in the UV domain is absorbed by intervening gas along the line-of-sight to QSOs.

The SF suppression at short $\Delta t$ has been observed across a range in luminosity: observations of low-luminosity AGN with the Kepler space mission have revealed breaks in the SF on time scales of days to weeks\cite{Mu11,Ka15a,Sm18}, although it has also been argued that there {\it ``remain significant levels of spacecraft-induced effects in the standard pipeline reduction of the Kepler data''} \cite{Ka15b}. Steeper slope changes have been seen in a recent study of luminous QSOs on time scales of 10 to 15 days\cite{St22}, while our break time scales range from $\sim 15$ to $\sim 50$ days. The reasons behind these breaks are still under debate: either the structure function is intrinsically different such as in the CHAR model\cite{Su20}, or a "smearing" effect may be caused by synchronised emission from different parts of the disk reaching the observer after different time lags\cite{Ta20}.

The forthcoming 10-year Legacy Survey in Space and Time (LSST \cite{Br18, Iv19, Ch20}) planned at the Vera C. Rubin Telescope for the years 2024-34 will eclipse this data set with better data: more passbands will probe a wider temperature range in the accretion disk. Deeper photometry will probe QSOs at much fainter luminosity. This, and the reduced photometric errors in LSST will provide low-noise structure functions for individual objects and reveal how tight and universal the structure function of QSO variability really is. LSST should also be a game changer for analysing subtleties in the short-term breaks in the SF.

\section*{Conclusion}

We have determined with unprecedented precision the parameters ($\gamma$, $B_L$, $B_\lambda$) of the variability structure function for the UV continuum emission of luminous QSOs. It is consistent with a universal random walk once time is expressed in units of the orbital or thermal timescales of the emitting material. The mean amplitudes of fluctuations can be described by the simple law $\log(A/A_0) = 1/2 \times \log (\Delta t/t_{\rm th})$, where a single fit parameter $A_0$ represents the overall QSO population. Note, that the thermal timescale has recently been shown to drive the damping timescale of QSO variability as well\cite{Bu21}, although this fact may not have affected the random walk at shorter timescales.

The result appears consistent with the temperature profiles of thin-disk models\cite{SS73}, at least at wavelengths from 130 to 300~nm. Deviations observed at longer wavelengths that show the cooler, outer parts of disks\cite{VB04, Mc10, Ca17, Yu22} still need to be explained. It also appears consistent with an origin of the fluctuations in magneto-rotational instabilities (MRI) in differentially rotating accretion disks that are made of magnetised plasma; there, the equations of motion scale with the orbital timescale\cite{Ba03,FKR02}. MRI is suggested to produce variability with a random walk phenomenology\cite{Kw98} and $\gamma \approx$ 0.4 to 0.5. Disks could thus be ordered in a 1-parameter family with the mean thermal timescale at a chosen wavelength as an ordering parameter. The structure function of QSOs may then be a universal relation apart from the short-term suppression. It appears intrinsically tight and would be broadened in observation by disk orientation and black hole spin, as well as host galaxy dust extinction affecting the inferred luminosity. It would also be broadened by intrinsic variety in vertical disk structure driven by accretion rates.

The role of the lamppost for driving QSO variability is likely confined to a contribution of modest amplitude on shorter timescales. If the CHAR model also described the physics of disk variability correctly, then it would be hard to disentangle genuine lamppost reverberation from outwards-propagating magnetic fields; and if the velocity of the latter dropped below the speed of light\cite{Su20}, the sizes of disks inferred by a reverberation analysis could be overestimated.

Future data from the 10-year LSST project will shed light on these questions and more subtle properties with its extremely precise measurements for a nearly all-sky QSO sample of unprecedented size, depth, variety and completeness.

\clearpage

\section*{Online Methods}

\subsection*{QSO sample and data cleaning}

We combined the Million Quasars Catalog \cite{Fl15} (MILLIQUAS v7.1 update) with the \textit{Gaia} eDR3 catalogue\cite{gaia21} and select 6,163 spectroscopically confirmed, non-lensed QSOs with Gaia magnitude $Rp<17.5$, redshift $0.5<z<3.5$, declination $\delta>-45^\circ$, Galactic foreground reddening\cite{SFD98} of $E(B-V)_{\rm SFD}<0.15$ and Gaia BpRp Excess Factor $<1.3$ (indicative of single, unblended, point sources). We also required that the MILLIQUAS and Gaia positions are coincident within $0.3''$, which excludes blended objects of similar SED such as the multiple images of lensed QSOs, and that sources have no Gaia neighbours within $15''$ to keep the ATLAS photometry unaffected. We confirmed with recent lists of lensed QSOs\cite{lenqso, Le22} that none of these are included in our sample. Our sample has 6,115 sources in common with NASA/ATLAS. The ATLAS photometric system is consistent with other reliable sources of photometry such as Pan-STARRS and Gaia \cite{To18b}.

The selection is based on Gaia eDR3 mean photometry, which is arguably the best-calibrated among large astronomical datasets\cite{Ri21}; the Gaia observations extend from mid-2014 to mid-2017 and overlap with the period of the ATLAS lightcurves (LCs). Thus, there will be no noticeable selection bias in favour of QSOs that might be brighter during an early selection epoch but fainter when the LCs are observed\cite{SB21}. 

We further remove six QSOs from the sample, which have large flux errors throughout their LCs (90-percentile flux errors in the orange band of $\sigma_{f_\nu, {\rm o}} > 150\mu$Jy or in the cyan band LC of $\sigma_{f_\nu, {\rm c}} > 85\mu$Jy). On individual images, the faintest objects in the sample have signal-to-noise ratios (SNR) of $>11$ in the orange band and $>13$ in the cyan band. Up to four images are available per band and night.

We reject radio-loud QSOs by cross-matching the sample with catalogues from the Faint Images of the Radio Sky at Twenty-cm \cite{Be95} (FIRST) Survey, the NRAO VLA Sky Survey \cite{Co98} (NVSS), and the Sydney University Molonglo Sky Survey \cite{Ma03} (SUMSS). We use matching radii between the MILLIQUAS and FIRST, NVSS, and SUMSS coordinates of 3", 12", and 11", respectively. 1,242 QSOs are matched with at least one radio catalogue. If a QSO is matched to multiple radio sources in NVSS or SUMSS, the radio detection with the closest separation is used. We apply a radio loudness criterion of $0.4\times(m_o-t_{\rm NVSS/SUMSS})>1.3$ for matches with NVSS and SUMSS. Radio sources below this threshold are labelled radio-intermediate. QSOs with matches in FIRST but not in NVSS turned out to be all radio-intermediate. The sample contains 775 radio-loud and 467 radio-intermediate objects, and 4,848 radio non-detections. Thus, our final sample contains 5,315 QSOs, of which 3,607 have single-epoch virial black hole mass measurements derived from SDSS spectra\cite{Ra20}.

We clean the ATLAS LCs first by excluding all observations with large errors of $\log(\sigma_{f_\nu, {\rm o}})>-3.94-0.12\times(m_{\rm o}-16.5)$ and $\log(\sigma_{f_\nu, {\rm c}})>-4.17-0.10\times(m_{\rm c}-16.5)$, which constitute about 5 percent of all observations; and further by comparing each measurement with other observations within $\pm$7 days and removing outliers with a $2\sigma$-clipping to reject spurious variability, e.g., due to temporary chance blending with faint asteroids. If there is only one observation within $\pm$7 days, it is retained.

\subsection*{Luminosity}

We estimate QSO luminosities at restframe 3,000 \AA\ ($L_{3000}$) using a linear interpolation of the Gaia $Bp$ and $Rp$ magnitudes. At redshifts of $0.5<z<0.7$ and $z>1.56$, the restframe 3,000 \AA\ point is extrapolated beyond the pivotal wavelengths of the Gaia filters. We correct for foreground dust extinction in the Milky Way\cite{SFD98} using bandpass coefficients\cite{CV18} of $R_{Bp}=3.378$ and $R_{Rp}=2.035$. Luminosity distances are derived in a flat $\Lambda$CDM cosmology with $\Omega_{\rm m}=0.3$ and a Hubble-Lema\^itre constant of $H_0=70$~km~sec$^{-1}$~Mpc$^{-1}$. Luminosity errors from this procedure are expected to be $<20$\% and thus much smaller than those introduced by host galaxy dust and the orientation of the disk. Luminosity estimates from broadband photometry does not distinguish between contributions from the accretion disk continuum and those from line emission, especially from the Fe complex that is strong in the vicinity of 3,000 \AA\ . Comparing our $L_{3000}$ estimates with those obtained by spectral fitting on the available subsample\cite{Ra20}, we find an average offset of $0.07$~dex and reduce our $L_{3000}$ values to correct for line contributions. Bolometric luminosities, as used by some authors \cite{Ca17}, are derived with $\log (L_{\rm bol}/L_{3000}) = \log (f_{\rm BC}\times \lambda) = \log (5.15\times 3000$\AA$) = 4.189$, using $f_{\rm BC}=5.15$ as the bolometric correction factor\cite{Ri06b}.

\subsection*{Thermal timescale}

For each QSO and passband, we estimate the characteristic (defined in the following) thermal timescale of the emitting material in the accretion disk. We combine the equation for $t_{\rm th}$ \cite{FKR02} and an equation for the scale length of the disk given $L,\lambda$ values\cite{Mo10}, and get:
\begin{equation}
    \begin{split}
	\frac{t_{\rm th}}{{\rm days}}=\frac{2.89}{1.8}\times\frac{1}{86400}\times10^{-13}\times(\frac{45GM_\odot f_{\rm BC} \lambda_{3000} L_{3000} \lambda^4_{\rm rf}}{16\pi^6\alpha^2 \eta h_p c^4 \cos i})^{0.5} \\
	=7.40\times10^{-27}\times(\frac{L_{3000}}{{\rm erg/s/\text{\AA}}})^{0.5}\times(\frac{\lambda_{\rm rf}}{\text{\AA}})^2  ~,
	\end{split}
	\label{eq:tth}
\end{equation}
where $h_p$ is the Planck constant, $\alpha=0.1$ \cite{Ki07} is the assumed viscosity parameter, and $i=45^\circ$ is the mean inclination. We assume a radiative efficiency of $\eta=L_{\rm bol}/(\dot{M}c^2)=0.1$, where $\dot{M}$ is the mass accretion rate; changes in $\eta$ due to variations in black-hole spin affect the inner cutoff of the accretion disk. For a given $\dot{M}$, such a variation has little effect on the outer disk, the overall $L_{3000}$, and the thermal timescales at the wavelengths observed here. Instead, it mostly affects the emission of higher-energy photons and $L_{\rm bol}$.

\subsection*{Variability structure function}

\begin{figure}
\begin{center}
\includegraphics[width=\columnwidth]{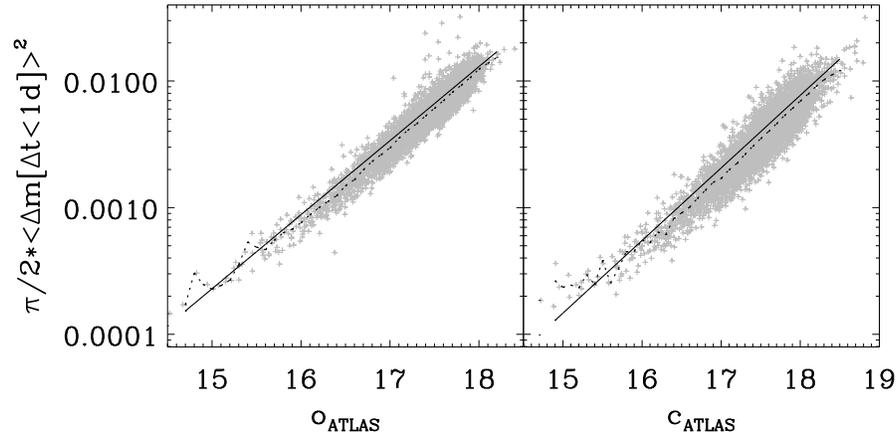}
\caption{The variability amplitude, $\frac{\pi}{2} <\Delta m>^2$, at $\Delta t < 1$ day as a function of observed magnitude. Gray symbols are individual QSOs. Dotted lines show the 3$\sigma$-clipped mean within fine magnitude bins. Solid lines show the fit of Equation~\ref{eq:noise}.
}\label{fig:noise_lv}
\end{center}
\end{figure}

We adopt the following definition of the variability structure function (SF)\cite{Cl96}:
\begin{equation}
	A=\sqrt{\frac{\pi}{2} <\Delta m>^2 - <\sigma^2>},   ~ ~ {\rm with} ~ ~
	\Delta m=|m_{\rm i}-m_{\rm j}|,
	\label{eq:sf}
\end{equation}
where $m_{\rm i}$ and $m_{\rm j}$ are any two observations and $\sigma$ is the magnitude error due to noise. We find that the magnitude errors quoted by ATLAS underestimate the noise and instead derive $\sigma^2$ from the apparent intra-day variability, which we expect to be negligible for radio-quiet QSOs. We determine $<\sigma^2>$ as a function of observed magnitude, $m_{\rm obs}$, such that the SF is anchored at $A\approx 0$ for $\Delta t < 1$ day. The required noise model is shown in Figure~\ref{fig:noise_lv} and given by 
\begin{equation}
	\log(<\sigma^2>)=n_0+n_1 m_{\rm obs} ~,
	\label{eq:noise}
\end{equation}
with $(n_0, n_1)=(-12.411, 0.585)$ and $(-12.428, 0.573)$ for the orange and cyan passband, respectively (pure Poisson noise suggests $n_1=0.4$). 

Informed by the pair sampling distribution of the LCs, we reduce window effects from the finite extent of the LCs by ignoring pairs with $\Delta t_{\rm obs} \geq 0.47 {\rm max}(\Delta t_{\rm obs})$. This cut is applied in the observed frame, while the rest of the paper uses the rest-frame $\Delta t = \Delta t_{\rm obs}/(1+z)$, corrected for cosmological time dilation.

\subsection*{Binning the time axis}

We bin the time axis in two different ways depending on purpose. A first case is used for the relation $\log A(\Delta t)$ in Equation~\ref{eq:SF_dt} and Figures~\ref{boot}, \ref{mean_parameters} and~\ref{SF_dt9}. A second case is used for fits over the final fitting range and for $\log A(\Delta t/t_{\rm th})$ in Equation~\ref{A_ttherm}, Figure~\ref{SF_dttherm} and Table~\ref{tab:sf_coes}.

In the first case, we split the $\Delta t$ axis into 20 bins. The first $\Delta t$ bin contains all pairs with $\Delta t < 1$ day, which is used only for noise analysis. Further bins are chosen to balance the number of pairs among the bins. The average number of pairs per $\Delta t$ bin are $\sim$103 million and $\sim$8 million for the orange and cyan passband, respectively. We split the $z$ axis into 9 bins, thus grouping QSOs with observations of similar restframe wavelengths. In the cyan passband, we exclude QSOs at $2.4<z<3.5$ to avoid contamination from the Ly-$\alpha$ line. In the orange passband, we use this z range as a single, highest-$z$, bin. The remaining eight bins cover the range of $0.5<z<2.4$ with roughly balanced QSO numbers. In each $z$ bin, QSOs are further split into ten $L_{3000}$ bins, each with almost equal QSOs numbers. In total, there are 1,800 bins in $(z,L_{3000},\Delta t)$. In each bin, the mean value of $z$ and $L_{3000}$ and the mid-point of $\Delta t$ is used for the analysis and figures.

In the second case, we split the $\Delta t$ axis into 100 bins. The shortest $\Delta t < 1$ day bin is exactly the same as in the first binning method, and further bins are balancing the number of pairs among the bins for both passbands. The QSOs are grouped into 16 bins along the $t_{\rm th}$ axis with an equal number of QSOs per bin, which is 2,950 and 4,134 in orange and cyan, respectively. In the $\Delta t/t_{\rm th}$ axis, we split the observed pairs into 100 bins of constant number, independently for each $t_{\rm th}$ group in each passband after excluding pairs with $\Delta t<1$ and $\Delta t>337$ days. Combining both passbands, there are in total 3200 bins for $(t_{\rm th}, \Delta t)$ and separately 3200 bins for $(t_{\rm th}, \Delta t/t_{\rm th})$. In each bin, the mid-point of $\Delta t$ and $\Delta t/t_{\rm th}$ is used for the analysis and figures.

\subsection*{Simple bin average structure function}

For each individual QSO, we calculate its SF with Equation~\ref{eq:sf} from all available pairs in each $\Delta t$ or $\Delta t/t_{\rm th}$ bin. We use a 3$\sigma$-clipped mean of magnitude differences as $<\Delta m>$. Then, for each $(z,L_{3000},\Delta t)$ or $(t_{\rm th}, \Delta t)$ or $(t_{\rm th}, \Delta t/t_{\rm th})$ bin we calculate the median squared amplitude, $A^2$, and the standard error of the median (SEMED), which ensures that QSOs with negative noise-corrected amplitudes $A^2$, due to an over-subtraction by the $\langle \sigma^2 \rangle$ noise model, are still included in the statistics.

\subsection*{Bootstrapping the structure function}

Using bootstrapping, we can account better for the contribution of each QSO. Here, we select random pairs in each of the $(z,L_{3000},\Delta t)$ or $(t_{\rm th}, \Delta t)$ or $(t_{\rm th}, \Delta t/t_{\rm th})$ bins. We randomly pick a QSO $N$ times from all the QSOs in that bin, where $N$ is the total number of QSOs in that bin. For each picked QSO, we randomly select 300 distinct pairs in the orange and 30 distinct pairs in the cyan passband, and then calculate the SF using Equation~\ref{eq:sf}. This process is repeated 100 times to get the bootstrap distribution, from which we calculate the mean and standard deviation for each bin.

\subsection*{SF fitting}

We fit a structure function of the form
\begin{equation}
	\log(A(\Delta t))=B_{0,\Delta t}+B_{L, \Delta t}\log(\frac{L_{3000}}{10^{42.7}{\rm erg}/{\rm s}/\text{\AA}})+B_{\lambda, \Delta t} \log(\frac{\lambda_{\rm rf}}{2200\text{\AA}}) 
	\label{eq:sf_l_lambda_fit}
\end{equation}
using a Levenberg-Marquardt least-squares fit\cite{Ma09} to the $(z,L_{3000})$ bins for each $\Delta t$ bin separately. Here, $\log A$ and its error may be results of bin averages or from the bootstrap method. $\lambda_{\rm rf}$ is the pivotal wavelength of a passband divided by $1+z$. The normalisation factors for $\lambda_{\rm rf}$ and $L_{3000}$ are chosen to be close to the median of the sample. Data with $\lambda_{\rm rf}>3000$\AA \ are excluded from the fit as it deviates from a linear relation. In Figure~\ref{SF_dt9}, we show an example for one $\Delta t$ bin: the left and middle panels show the two passbands. The right panel shows the intercepts of the fits at $\log(L_{3000})-42.7$ for different $\lambda_{\rm rf}$. We also compare the upturn at longer wavelengths with results from the literature\cite{Ca17}. The $B_{0,\Delta t}$, $B_{L, \Delta t}$, and $B_{\lambda, \Delta t}$ from each $\Delta t$ bin are shown in Figure~\ref{boot}.

\subsection*{Global parameter estimates using broad $\Delta t$ ranges}

We fit a global model of the form
\begin{equation}
	\log(A)=B_0+B_{\rm L}\log(\frac{L_{3000}}{10^{42.7}{\rm erg}/{\rm s}/\text{\AA}})+B_\lambda\log(\frac{\lambda_{\rm rf}}{2200\text{\AA}})+\gamma\log(\frac{\Delta{\rm t}}{216{\rm d}})
	\label{eq:sf_coes_dt_fit}
\end{equation}
to the SF data in the $(z,L_{3000}, \Delta t)$ bins with $\lambda_{\rm rf} <$ 3000\AA \. We test the robustness of global parameter averages by varying the time span of data that is included in the global fit; we vary the lower edge $\Delta t_0$, while leaving the upper edge at $\Delta t$=337~days unchanged. Figure~\ref{mean_parameters} shows the parameter estimates as $f(\Delta t_0)$.

\subsection*{The structure function as $f(\Delta t)$ and $f(\Delta t/t_{\rm th})$}

\begin{figure}
\begin{center}
\includegraphics[width=\columnwidth]{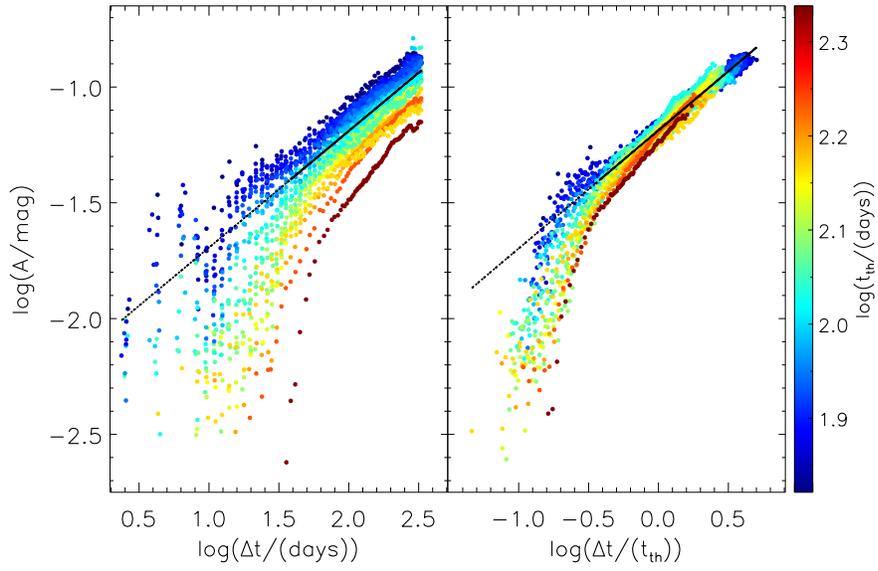}
\caption{Variability amplitudes $\log A$ vs. $\log \Delta t$ (left) and vs. $\log (\Delta t/t_{\rm th})$ (right). The colour encodes the thermal timescale of the measurement from blue for the shortest to red for the longest $t_{\rm th}$. The best-fit lines to the data at $\log A > -1.4$ have slopes of $\gamma=0.503$ (left) and $\gamma_{\rm th}=0.510$ (right), consistent with a random walk. Solid lines show the fit range while the dotted lines are extrapolated. At short $\Delta t$, the variability appears suppressed and this effect reaches to longer $\Delta t$ in larger (here: redder) disks.
}
\label{SF_dttherm}
\end{center}
\end{figure}

For Figure~\ref{SF_dttherm} we split the sample for each passband separately into 16 groups ranked by $t_{\rm th}$. We then fit Equation~\ref{eq:sf_coes_dt_fit} to the points with $\log A>-1.4$ using the errors for weighting (see Table~\ref{tab:sf_coes} and left panel of Figure~\ref{SF_dttherm} for results). Similarly, we fit the bins with $\log A > -1.4$ to Equation~\ref{A_ttherm} (see Table~\ref{tab:sf_coes} and right panel of Figure~\ref{SF_dttherm} for results). For reference, we find $\log A_0 = -1.185$ when fixing $\gamma=1/2$ in the bootstrap method.

\section*{Data availability}
The QSO samples are selected from the publicly available MILLIQUAS database complemented with publicly available Gaia data. Data from NASA/ATLAS are publicly available at \href{https://fallingstar-data.com/forcedphot/}{https://fallingstar-data.com/forcedphot/}. The QSO light curves used in this study are available from the corresponding author on request.

\section*{Code availability}
The main analysis routines have been written by the corresponding author in IDL and are available on request.

\section*{Acknowledgements}
JJT was supported by the Taiwan Australian National University PhD scholarship and the Australian Research Council (ARC) through Discovery Project DP190100252 and also acknowledges support by the Institute of Astronomy and Astrophysics, Academia Sinica (ASIAA). JT has been funded in part by the Stromlo Distinguished Visitor Program at RSAA. We thank Mark Krumholz for comments on the manuscript and Stefan Wagner for helpful discussions.

This work uses data from the University of Hawaii's ATLAS project, funded through NASA grants NN12AR55G, 80NSSC18K0284, and 80NSSC18K1575, with contributions from the Queen's University Belfast, STScI, the South African Astronomical Observatory, and the Millennium Institute of Astrophysics, Chile.

This work has made use of SDSS spectroscopic data. Funding for the Sloan Digital Sky Survey IV has been provided by the Alfred P. Sloan Foundation, the U.S. Department of Energy Office of Science, and the Participating Institutions. SDSS-IV acknowledges support and resources from the Center for High Performance Computing  at the University of Utah. The SDSS website is \href{http://www.sdss.org}{http://www.sdss.org}. SDSS-IV is managed by the Astrophysical Research Consortium for the Participating Institutions of the SDSS Collaboration including the Brazilian Participation Group, the Carnegie Institution for Science, Carnegie Mellon University, Center for Astrophysics | Harvard \& Smithsonian, the Chilean Participation Group, the French Participation Group, Instituto de Astrof\'isica de Canarias, The Johns Hopkins University, Kavli Institute for the Physics and Mathematics of the Universe (IPMU) / University of Tokyo, the Korean Participation Group, Lawrence Berkeley National Laboratory, Leibniz Institut f\"ur Astrophysik Potsdam (AIP),  Max-Planck-Institut f\"ur Astronomie (MPIA Heidelberg), Max-Planck-Institut f\"ur Astrophysik (MPA Garching), Max-Planck-Institut f\"ur Extraterrestrische Physik (MPE), National Astronomical Observatories of China, New Mexico State University, New York University, University of Notre Dame, Observat\'ario Nacional / MCTI, The Ohio State University, Pennsylvania State University, Shanghai Astronomical Observatory, United Kingdom Participation Group, Universidad Nacional Aut\'onoma de M\'exico, University of Arizona, University of Colorado Boulder, University of Oxford, University of Portsmouth, University of Utah, University of Virginia, University of Washington, University of Wisconsin, Vanderbilt University, and Yale University.

This research has made use of data obtained from LAMOST quasar survey. LAMOST is a National Major Scientific Project built by the Chinese Academy of Sciences. Funding for the project has been provided by the National Development and Reform Commission. LAMOST is operated and managed by the National Astronomical Observatories, Chinese Academy of Sciences.

This work has made use of data from the European Space Agency (ESA) mission Gaia (\href{https://www.cosmos.esa.int/gaia}{https://www.cosmos.esa.int/gaia}), processed by the Gaia Data Processing and Analysis Consortium (DPAC,\href{ https://www.cosmos.esa.int/web/gaia/dpac/consortium}{ https://www.cosmos.esa.int/web/gaia/dpac/consortium}). Funding for the DPAC has been provided by national institutions, in particular the institutions participating in the Gaia Multilateral Agreement.

\section*{Author Contributions Statement}
JJT contributed the data analysis and drafting and editing of the article. CW contributed the conception and design of the work, the supervision of JJT, as well as drafting and editing of the article. JT contributed the observation and data collection.

\section*{Competing Interests Statement}
The authors declare no competing interests.

\clearpage

\begin{table}
	\begin{center}
	\caption{Coefficients for final fit results. }
	\label{tab:sf_coes}
	\begin{tabular}{lcc}
		\hline
		method                  & bootstrap             & bin averages          \\
		\hline
		$B_0$		            & $-0.944\pm0.001$		& $-0.985\pm0.001$		\\
		$B_L$		            & $-0.271\pm0.002$		& $-0.277\pm0.004$		\\
  		$B_\lambda$		        & $-1.216\pm0.011$		& $-1.265\pm0.021$		\\
		$\gamma$ 		        & $+0.503\pm0.002$		& $+0.447\pm0.003$		\\
		$C_L$ 		            & $+0.539\pm0.004$		& $+0.620\pm0.010$		\\
  		$C_\lambda$ 	        & $+2.418\pm0.023$		& $+2.831\pm0.052$		\\
		\hline
		$\log A_0$              & $-1.187\pm0.001$      & $-1.210\pm0.001$      \\
  		$\gamma_{\rm th}$       & $+0.510\pm0.002$      & $+0.463\pm0.003$      \\
		\hline
	\end{tabular}
	\end{center}
\end{table}

\clearpage

\end{document}